\documentstyle[prl,aps,amssymb,graphicx]{revtex}
\newcommand{{ \vn }}{\vec{n}}

\newcommand{\beq}{\begin{equation}}
\newcommand{\eneq}{\end{equation}}

\newcommand{\met }{\frac{1}{2}}
\newcommand{\Seff }{S_{\text{\it eff}}}
\newcommand{\sa }{\sigma_{\alpha}}
\newcommand{\sab }{\sigma_{\bar {\alpha}}}
\newcommand{\ab }{\bar {\alpha}}
                                             \tighten
                                              
\begin{document}
                                             \draft
           \twocolumn[
          \hsize\textwidth\columnwidth\hsize\csname@twocolumnfalse\endcsname
           \title{Kondo ground state in a quantum dot with 
  an  even number of electrons    in a  magnetic field}
            \author {  D. Giuliano$^{1,2}$  B.Jouault$^{1,3}$ and 
             A. Tagliacozzo$^{1,4}$}
        \address{ $^{1}$Istituto Nazionale di Fisica della Materia (INFM),
         Unit\'a di Napoli\\
         $^{2}$ Department of Physics, Stanford University,
         Stanford, California 94305\\
         $^{3}$ GES, UMR 5650, Universit\'e Montpellier $II$,
         34095 Montpellier Cedex 5, France\\
         $^{4}$ Dipartimento di Scienze Fisiche Universit\`a di Napoli
         "Federico II ",\\
         Monte S.Angelo - via Cintia, I-80126 Napoli, Italy}
         \date{\today}
         \maketitle
\begin{abstract}
Kondo conduction has been observed in a quantum dot with an even number 
of electrons at the Triplet-Singlet  degeneracy point produced by applying 
a small magnetic field $B$ orthogonal to the dot plane. 
  At a much larger  field
$ B \sim B_*$,  orbital effects induce  the reversed transition from the 
Singlet to  the  Triplet  state.  We study the  newly  proposed  Kondo 
behavior  at this point.
Here   the Zeeman spin splitting cannot be  neglected, 
what  changes   the nature of the Kondo coupling. On grounds of exact 
diagonalization  results in a dot with  cylindrical symmetry, we show that, 
at  odds with what happens at the other crossing point, close to $B_*$,
 orbital and spin degrees of freedom are ``locked together'', so 
that  the Kondo coupling involves a fictitious  spin $1/2$ only, which 
is fully compensated by conduction  electrons under suitable conditions. 
In this sense, spin at the dot is  fractionalized. 
We derive the scaling equation of the system by means of
a nonperturbative variational approach. The approach is extended to the
$B \neq B_*$-case and  the residual magnetization
on the dot is discussed.

\end{abstract}

\pacs{
\hspace{1.9cm}
}
]

\section{Introduction}

Low temperature transport properties of quantum dots (QD) are
primarily determined by electron-electron  interactions\cite{kouwenhoven}. 
When a QD is weakly coupled
to the contacts its charging energy is larger than the thermal energy. 
Therefore, in the absence of a  source-drain voltage bias, $V_{sd}$,
electrons can be added to the dot in a controlled way by changing 
an applied  gate voltage, $V_g$ \cite{tarucha0,schmid}.
 Conductance peaks are separated by 
 Coulomb Blockade (CB) regions, in which the electron number $N$ is fixed
(Coulomb oscillations). 

However, there is a well established  experimental evidence 
for   zero-bias anomaly 
in the conductance in a  CB region   when  $N$ is odd 
\cite{goldhaber,cronenwett}. 
This occurs when  the temperature is further decreased 
and  the strength of the coupling to the contacts increased. 
Such a behavior is attributed  to the formation of a Kondo-like hybridized 
state between the QD and the 
contacts \cite{theo,hewson}. 
 Kondo effect is not expected when $N$ is even,  because of 
lack of spin degeneracy as the QD is in a singlet state. 
An exception  is provided  when a QD with  
$N$ even is in a triplet state because of  Hund's rule. In this case the
QD shows a very peculiar spin-1 Kondo Effect 
\cite{silvano,tarucha,eto,pg}.  The Kondo anomaly  is enhanced if 
the singlet state is close to be degenerate with the triplet. 
Such a regime can be achieved experimentally by 
applying a weak magnetic field  orthogonal to the dot,
  $B_\perp $, because Hund's rule breaks down quite soon. Correspondingly, 
the QD undergoes a triplet-singlet
transition, which eventually suppresses the Kondo effect. In the following 
we shall refer to the transition above as ``TS crossing''.

If the magnetic field is further increased, 
orbital effects induce the reverse
transition from the singlet to the triplet state at some ``critical
field'' $B_\perp  = B_*$, in the general landscape of 
transitions to higher spin states \cite{wagner,jouault}. In this paper
we focus on  this transition, in the following referred to as ``ST crossing''.
 The accidental degeneracy between states with different total spin
can restore the Kondo effect. However, as we will show below, the presence
of a substantial Zeeman splin splitting (Zss) makes the ST crossing
very different from  the TS crossing.
 Our analysis is an extension of the results of \cite{noi}.    

We consider a vertical QD with contacts located at the top and at 
the bottom  of a pillar structure \cite{tarucha0} 
(see Fig.~\ref{scheme}a).  Because the confining potential is chosen to be 
parabolic in  the  radial direction, single  particle states on the 
dot are labeled by  $(n,m,\sigma)$ where $n$ is an integer, 
$m$ is the  orbital angular  momentum  and $\sigma$ the electron spin. 
  We have  chosen  $N=2$ for sake of simplicity, but 
we believe that the pattern we describe is quite frequent. For example, it
 can be extended to the case when $N$=4.  

Because of the Zss, the crossing occurring at  the ST point 
involves  the singlet and the $S^z=1$-component of the triplet  state
only.
 The analysis of the quantum numbers of the  $N\pm 1$ 
states visited  by virtual transitions  (cotunneling processes )  shows 
that orbital and spin quantum numbers of the electrons involved are
``locked together'' (Fig. 2a). Tunneling from the contacts does not conserve 
$n$, but, since the geometry is cylindrical, it conserves $m$ and the spin.
In particular, if the dot is in the triplet state, only  
$ (m=0, \downarrow )$-electrons can enter it, while is it is in the singlet
state, then only $(m=1, \uparrow )$-electrons enter.
 This envisages a one-channel  spin 1/2-like Kondo coupling different from
 what  occurs at the TS point. 

 The existence of a Kondo
effect between states belonging to different representations of the total
spin is very peculiar. Being the total dot spin of both the degenerate levels 
integer  ($S=0,1$), spin compensation has to be incomplete. 
In this work we show that it is as if the total spin were  decomposed into 
two  fictitious   spins $\met$ which we refer to as $S_{r}$ 
and $\Seff $ \cite{nota2}. We stress that these dynamical variables 
involve both orbital and spin degrees of freedom which are locked together. 
 The magnetic field $B_*$  favors antiferromagnetic (AF) coupling between 
the delocalized electrons of the contacts and $\Seff $, which quenches it 
in the strong coupling limit. $S_r$ is the residual spin $\met $ at
the  dot\cite{nota3}.     
CB pins $N$ to an even value, but the correlated state that sets in will
have half-odd spin, i.e., ``fractionalization of the spin'' at the QD site
will take place \cite{noi}. $B_*$  only affects the magnetization related to 
$S_r$, but we will not discuss such a point in this paper. On the contrary,
detuning $B$ from $B_*$ corresponds to an effective magnetic field
$\delta B = B - B_*$, which affects Kondo screening.

On the other hand, at temperatures $T > T_K$ ( $T_K$ being the Kondo 
temperature), there is no Kondo correlation.
Furthermore, being the dot in CB, the screening by the 
delocalized  electrons is quite negligible. As long as 
$B < B_*$,  the total spin is 0. For  $B>B_*$, the 
total spin is 1.
The spin density on the dot has a drastic change 
at $B_*$, while  the charge density is only slightly 
affected.  This is shown in Fig. 3a: for $B < B_*$ (singlet state) 
the spin density 
is zero, while it is half the charge density for $B > B_*$ (triplet state 
$S=1$ with $z-$ component  $S_z =1$).

This shows that the total spin of the dot is the relevant dynamical variable 
in the correlation process that sets in when temperature is lowered,
while the charge degree of freedom is frozen for $B \sim B_*$. 

In Section IV  we set up  a variational technique similar in spirit to 
Yosida's approach \cite{yosida} for the  construction of 
 a trial wavefunction  of the correlated Ground State (GS). 
We start from a Fermi sea (FS) of delocalized electrons of the 
contacts, times the impurity spin wavefunction and we  project out 
states which do not form a singlet at  the dot  site.
This is  a modified version of the Gutzwiller projector and  
the results we obtain qualitatively reproduce  the strongly-coupled
regime of the Kondo model \cite{hewson,iaff,tswieg,compens}. 
Our formalism also allows us to discuss  
the $\delta B \neq 0$-case  and 
 to calculate ``off-critical'' quantities, like the 
magnetization, as a function of $\delta B$.
In \cite{noi}  we estimated the Kondo  temperature $T_K$. 
It is slightly below the border of 
what is reachable nowadays within a transport measurement 
($T_K \sim$ tens  of  mK). Instead, an  Electron 
Paramagnetic Resonance (EPR) 
experiment could provide  evidence for the 
fractionalization of the spin. 
At $T<T_K$, the system goes across the
Kondo state when $B$ is moved across  $B_*$.
Partial compensation of the dot spin due to  the local screening  
implied by   the Abrikosov-Suhl resonance takes place at the QD site.
 The residual magnetization  is the one of  a spin-1/2 in all the range 
$B \sim B_*$.   
Therefore, in lowering the temperature, the response to the microwave field 
 should also 
appear all the way from  $B < B_*$ to  $B >B_*$  and 
a temperature-dependent Knight
shift of  the  frequency should occur.
Moreover,  this feature should be  detectable, due to the enhanced 
susceptibility of the Kondo state.

Differently from the scenario described  here, the TS crossing has 
 a nonsymmetric behavior  across  the degeneracy point.
Spin 1 Kondo effect should take place for $B$ prior to the degeneracy point, 
It is renforced at the TS transition, but  no Kondo effect at all 
should occur when the GS is a singlet.     

The paper is organized as follows:

\begin{itemize}

\item  In Section II we show that $B_*$
makes only four levels involved in the Kondo coupling, in analogy 
to the single impurity  spin $\met$  Anderson
model (AM). We discuss the strongly coupled regime of the system by briefly 
recalling the scaling theory of the AM  and specify what we mean as
spin fractionalization.

\item In Section III we employ the Schrieffer-Wolff transformation to
map our system onto an Effective Kondo model. We show that the
corresponding
interaction between $\Seff$ and the spin of the electrons
from the Fermi sea is antiferromagnetic (AF), which generates the 
strongly coupled Kondo regime.

\item In Section IV we construct a trial GS using a modified
version
of the Gutzwiller projection method. We calculate the
energy, which depends on the  exchange interaction: 
$j$. We show that $j$  scales
according to  Anderson's poor man's scaling \cite{poorman}. 

\item In Section V we extend our technique to the $\delta B \neq 0$-case.
 Because of  the presence of $\delta B$,  
a  second scaling parameter, $h$, arises. 
We compute the energy and 
the magnetization  as a function of $h$ and discuss   how this is 
related  to the bare field $\delta B$.

\item  In Section VI we report our final remarks and conclusions.
\end{itemize}

\section{The model for the dot at the level crossing ($B=B_*$).}

\begin{figure}
\centering \includegraphics*[width=0.9\linewidth]{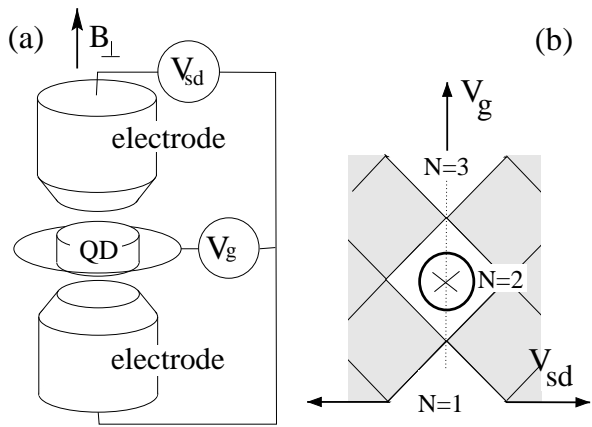}
\caption{ (a) Vertical Quantum Dot with contacts in a pillar configuration.
(b) Grey-scale plot of the differential conductance $vs$ gate 
voltage $V_g$ and biasing voltage $V_{sd}$ (schematic), showing the CB 
regions (white) at various $N$.  Operation point is marked by $\otimes $.}
\label{scheme}
\end{figure}

Using exact diagonalization\cite{jouault}, two of us studied few interacting 
electrons in 
two dimensions, confined by a parabolic potential 
in the presence of a
 magnetic field $B$ orthogonal to the dot disk  ($z-$direction).
 We chose 
the energy separation between  the levels of the confining potential 
  $ \hbar\omega _d$ to be of the same size as  the effective strength of the  
Coulomb repulsion $U$ ($\sim 4 meV$), 
which makes the correlations among the electrons relevant.  
The magnetic field   favors the increase of the total
angular momentum in the $z-$ direction, $M$,  as well as of the total spin  
of the dot, $S$, in order  to minimize the Coulomb repulsion.
The $B$-dependence of the spectrum of the QD energy levels 
can be monitored via a linear transport measurement, by attaching 
electrodes to it in a pillar structure  (Fig.~\ref{scheme}) 
\cite{goldhaber,cronenwett,tarucha}.
At special values $B=B_*$, one can identify several level crossings 
between states with different $M$ and $S$.

In this paper we shall focus onto a QD
with $N=2$ electrons, but the features we discuss here are appropriate 
for dots with any even $N$. 
The parameters of the device can be chosen 
such that 
 $B_*$  is rather large, which lifts the spin degeneracy, because of  
Zeeman spin splitting \cite{nota}. We define $B_*$ as the field at which  
the GS  of the isolated dot switches from singlet 
($~^2 S^0_0$; $S=0, M=0$) to  the  triplet state lowest in energy
(the state  $S_z =1$ of  the three states $~^2 T^1_{0,\pm 1}$: 
$S=1, M=1$ and $ S^z =0,\pm 1$).

\begin{figure}
\includegraphics*[width=0.9\linewidth]{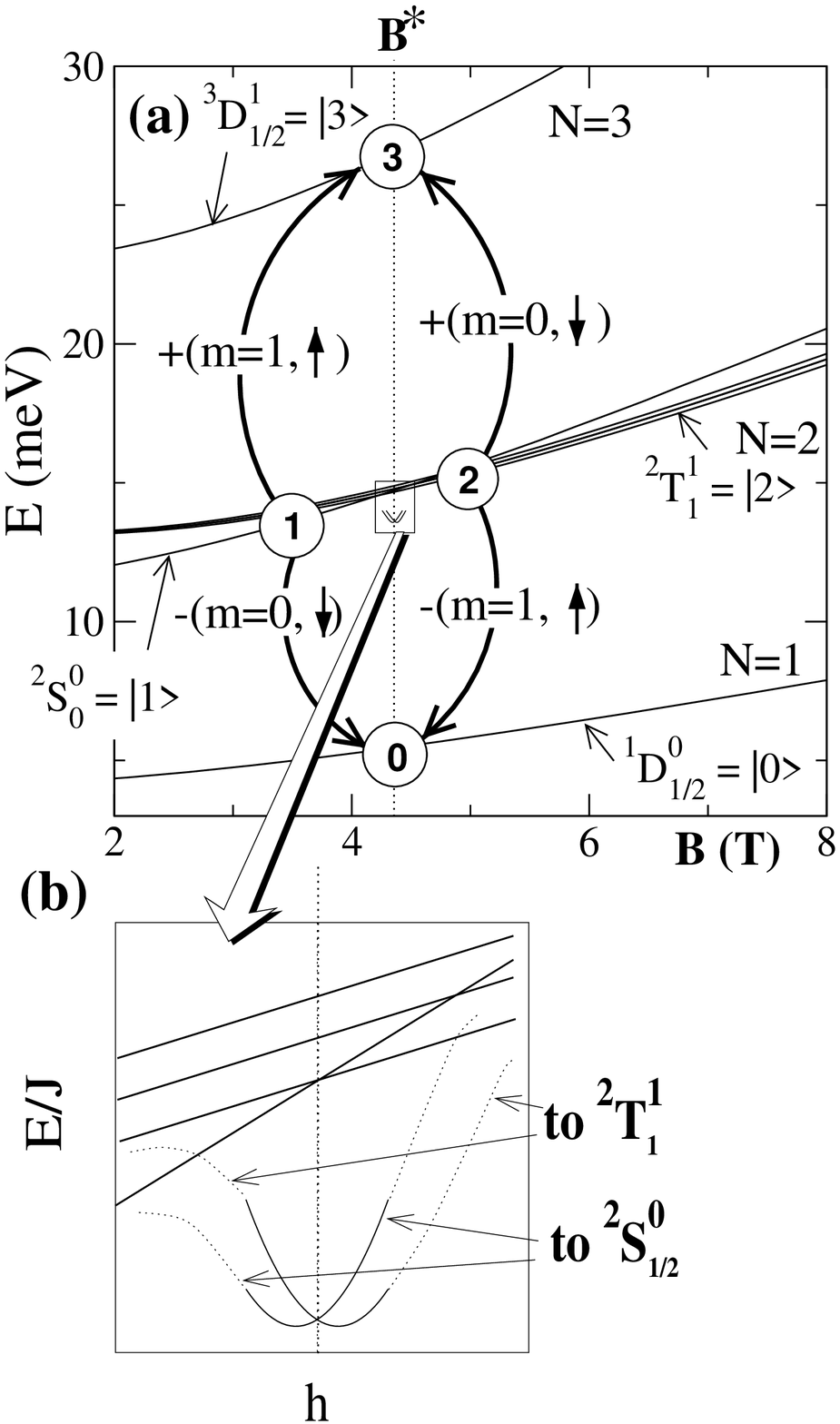}
\caption{(a) General scheme of the energy level of the QD $vs$ magnetic 
field $B$. Only few states are plotted (labels are defined in Section II).
 The level spacing of the parabolic lateral confinement of the QD 
is $\hbar \omega_0$= 4meV , the Coulomb 
interaction $U$ = 3meV. Circled numbers represent the four dot states 
involved in the Kondo resonance. (b) 
Energy gain of the Kondo correlated state (not in scale) and removal 
of the degeneracy at $\delta B =B-B_* \neq
0$ ( $ h \propto \delta B / T_K$: see Section V).}
\label{kondo}
\end{figure}

Only two $N=2$ states  are primarily involved 
at the level crossing, namely, the $S=0$ singlet and the $S=1, M=1, 
S^z = 1$ component of the triplet (see Fig.~\ref{kondo}a). At the same $B$, the
 low lying $N=1$ and $N=3$-GS's are spin doublets. We shall denote 
them as $~^1 D_\frac{1}{2}^0$  and  $~^3 D_\frac{1}{2}^1$, respectively. 
Lifting the spin degeneracy 
makes the $S^z = \frac{1}{2}$ components lower in energy in both cases. If
 $V_g$ is tuned in the CB region between $N=2$ and $N=3$ (Fig.~\ref{scheme}b), only
the four states listed above are primarily relevant for the conduction.

We now construct a model  by
taking  the 1-particle state , $~^1 D_\frac{1}{2}^0$,  as the ``vacuum'' 
of the truncated Hilbert space $\Xi$ (which  we denote  
by $ | 0 \rangle$). 
Let us take  $ | 0 \rangle$  to be the initial state of the QD.
The dot can exchange
one electron at a time with the contacts. A $(q,m,\uparrow) $-electron 
entering the QD  generates a transition between $ | 0 \rangle$ and $~^2 T^1_1$.
Alternatively, the  lowest energy state  available 
 for a $(q,m,\downarrow )$-electron entering the QD  is  the singlet
$~^2 S_0^0$ (here $q$ is the modulus of the momentum of the particles and $m$
is the azimuthal quantum number, which is conserved by the interaction with the
dot). It is useful to  define the operators $d_{ 1,2}^\dagger$ which
create the two $N=2$ many body states $ |\alpha \rangle $
 ($\alpha =1,2 )$, belonging to $\Xi$,  by acting on $| 0 \rangle$:
\beq
|1\rangle \equiv | ~^2 S^0 _0\rangle = d_{ 1}^\dagger | 0 \rangle ; \: \:
\:|2\rangle \equiv
| ~^2 T^1_1 \rangle = d_{ 2}^\dagger | 0 \rangle
\label{states}
\eneq
 The operator 
$d_{ 1 }^\dagger  /d_{ 2 }^\dagger $ adds a  $(m=0 ,\downarrow ) /
(m=1 ,\uparrow  )$ electron to the dot.
Finally, the state $ |3\rangle $ is defined as: $ |3\rangle 
  \equiv | ~^3D_{\met}^1 \rangle = 
d_{ 2}^\dagger d_{ 1}^\dagger | 0 \rangle $.
The operators   $d_{1,2}$ defined above cannot be associated to 
any  single particle function  on the dot. Nevertheless,
as far as low-energy excitations only are involved, it is a good approximation
to take them to be anticommuting.
To show this we refer to  Fig.~\ref{weight},  where we report the calculated 
quasiparticle spectral  
 weight, when   an extra electron  is added to 
  the dot  in one of the   two 
possible GS's  at $B = B_*$  and  $N=2$.  In Fig.~\ref{weight}a we show 
the weight for  addition   of an extra  particle to the state 
$ | ~^2 T^1_1 \rangle $ to give a state with 
total angular momentum   $M=1$.  This is close to 1  for a 
$\downarrow $-spin but is practically zero at low energies for an
$\uparrow $ spin because  states with $S=3/2$ are higher in energies.  
According to the definitions of eq.(\ref{states}), this implies that, 
while the state 
$ d_{ 1}^\dagger d_{ 2}^\dagger | 0 \rangle $ can be normalized,  one has 
$ (d_{ 2}^\dagger)^2 | 0 \rangle \approx 0 $.  
Analogously, Fig.~\ref{weight}b  shows that  there is an energy  shift  
of the  spectral weight peak
 when a $\uparrow $ or a $\downarrow $ spin particle  is added  
to the state $ | ~^2 S^0_0 \rangle $. This shift corresponds to  the Zeeman 
spin splitting between the states $^3 D ^1_{\met}$ and  $^3 D ^1_{- \met}$. 
 We ignore  the latter that  is placed at higher energy. In  
 Section  III we argue that inclusion  of this state does not 
qualitatively change  our results. 
Then, within this approximation that includes the 
lowest energy state only, we take   
$ (d_{1 }^\dagger)^2 | 0 \rangle \approx 0 $. So, we are alleged to assume  
$ ( d_{ 1,2 }^\dagger )^2 = 0$. Hence,
 the $d$-operators behave as they were single-particle 
 fermion operators, although, in fact, they create many-particle states. 
 
We write  the  Hamiltonian for the QD as:
\beq
H_D = \epsilon_d ( n_{1} + n_{2} ) + U n_{1} n_{2}
\label{doth}
\eneq

where $n_{\alpha} = d_{ \alpha}^\dagger d_{  \alpha}$,
$\epsilon_d$ is the energy of the degenerate $N=2$ levels and $U$ is
the charging energy for adding a third electron.

We assume the electrons in the leads to be noninteracting. 
Let  $ b_{ j q \sigma}$ be the annihilation operators
  for one  electron in 
the left ($j=L$) and the right ($j=R$) contact, respectively, and let 
$\epsilon_q$ be the corresponding energy (independent of $j$ and 
$\sigma$ for simplicity).
The Hamiltonian for the leads is:

\beq
H_{l} = \sum _{ q  \sigma}\sum _{ j = L,R} 
\epsilon _q  b_{ j q \sigma}^\dagger  b_{ j q \sigma } 
\label{leadh}
\eneq
We denote by ${\cal{H}}_0 = H_D +H_l $ the sum of the two Hamiltonian terms
given by eqs.(\ref{doth},\ref{leadh}).  They do not change the 
particle number $N$ on the dot. Instead, $N$ is changed by 
 tunneling  between the leads  and the dot, according to the 
above mentioned selection rules.
Let $\Gamma_R$ and $\Gamma_L$ be the  amplitudes for single-electron
 tunneling
 between the dot and the $R$ or  $L$ contact, respectively. Removal of one 
electron from the contacts is described by the following linear combination 
of the $ b_{ j q \sigma}$'s:

\beq
\phi_{ q \sigma} \equiv \frac{ 1}{ \sqrt{|\Gamma _L|^2 +|\Gamma _R |^2}} 
( \Gamma _L b_{ L q \sigma } +
\Gamma _R b_{ R q \sigma } )
\eneq 

For a point-like QD, there will be no tunneling term 
involving  the  combination of $ b_{ j k \sigma}$ operators    
linearly independent of  $\phi_{  q \sigma}$.
The  terms to be added are $(\alpha = 1,2)$:
\beq
V_{\alpha 0} = \sum_q v_{\alpha 0}^* ( q ) d_\alpha ^\dagger 
\phi_{ q \sigma_{\alpha}}
\label{ofdia1}
\eneq
for $N=1 \to N=2 $ and:
\beq 
V_{3\alpha } = \sum_q v_{3 \alpha } ( q ) d_{\bar {\alpha }}^\dagger 
\phi_{ q \sigma _{\bar {\alpha}} },
\label{ofdia2}
\eneq
for  $N=2 \to N=3  (\ab \neq \alpha)$ . Because spin is conserved in the 
tunneling  is  $\sigma _1 = \downarrow$, and 
$\sigma _2 = \uparrow$ and   $\bar{\alpha} \neq \alpha $. 
 Hermitian conjugate operators of 
$V_{\alpha 0}$ and $ V_{3 \alpha }$ decrease the 
electron number on the dot.  

This  model  Hamiltonian  has  the form of   
the nondegenerate Anderson Model (AM).
In \cite{noi} we have considered a special choice of the $V_g $ 
fine tuning  (see Fig.~\ref{scheme}b), which corresponds to  
the symmetric case:
the singly occupied localized impurity level has energy   $\epsilon _d
= - U/2$, 
while the doubly occupied one has energy  $\epsilon_d +U = U/2$.
  Also, $\Gamma _j\:\: (j=L,R)$ have  been taken independent on $j$.

Integration over the contact fields  
$\phi _{q\sigma}$  in the 
partition function leads to the   constrained
dynamics of the dot  between the two degenerate states $|\alpha \rangle$, 
labeled by $\alpha =1,2$ and coupled to an  external fluctuating 
field  $X(\tau)$ \cite{hamann}. As shown in \cite{noi},
 the main contributions to the 
partition function  are the quantum fluctuations of an effective 
fictitious spin $\met$
$\Seff$, due to the coupling among the dot and the leads.  
This spin flips from 
1 to 2 because of cotunneling processes  in which an $( m=1,\uparrow )$ 
electron is added/removed in the 
dot while an $(m =0,\downarrow )$ electron is removed/added 
(see Fig. 2a )  Its 
quantum dynamics  produces a correlated state between dot and leads 
in which the  charge excitations, being higher in energy, are fully 
decoupled  and suppressed. 
Four states are available prior to correlation, corresponding to $S = 0,1 $, 
two of which  are degenerate. 
They factorize according to the decomposition law:
\beq
\vec{S} = \vec{S_r} + \vec{S}_{\text{\it{eff}}}.
\label{phef}
\eneq
where $S_r$ is a residual spin $1/2$ whose wavefunction 
 factorizes  close to   $B_* $.  Correlation with the contacts  is governed by 
  a single channel spin $\met $ AF Kondo Hamiltonian 
for $\Seff$, which is 
derived  in Section III and discussed in detail in Appendix A. In the 
corresponding GS, $\Seff$ is fully compensated as in the usual Kondo Effect
at zero magnetic field.

In Sect. III and IV  we show that, provided that the coupling  to the
leads is symmetrical for $\alpha$=1,2, the GS is doubly degenerate  at $
B= B_* $ (the limiting states are derived from the  $|g^\lambda \rangle$ 
$(\lambda = \pm )$ of eq.(\ref{trial1}), having energy $\epsilon (\xi ,j ) $ 
given by eq.(\ref{ener2})).
The degeneracy disappears  when is $\delta B = B - B_* \neq 0$ (see Fig.2b). 
 Such a spin fractionalization closely resembles quantum number 
fractionalization  in strongly-correlated
1-dimensional electron systems \cite{fadeev}.  

\section{Effective AF  Kondo interaction.}

In the previous Section we have introduced the four states $ | i \rangle$,
$i = 0,1,2,3$ ,  which are primarily relevant to the low-temperature dynamics 
of the QD  (see eq.(\ref{states})). Tuning  $V_g$ as described in
Section II (see Fig.~\ref{scheme}b) allows only the $N=2$ states, $|1 \rangle$ and $ | 2 \rangle$, 
to be the initial and final states of any process. Let $\Xi$ be the subspace
they span. The $N  \neq 2$-states, $| 0 \rangle$ and $ | 3 \rangle$,
play a role only as virtual
states allowing for higher-order tunneling processes. 
Then, it is desirable  to
describe the dynamics of the dot in terms of an effective Hamiltonian
acting on  $\Xi$ only.
This is accomplished in this Section, by  employing   the 
Schrieffer-Wolff (SW) transformation  at second order
in the tunneling amplitudes.

According to eqs.(\ref{doth},\ref{leadh} ) the elements of  ${\cal H}_0 $
 are diagonal in the Fock space  and are given by:  
\beq
H_{00}=H_l ;\; H_{ 11} = H_{22} = H_l + \epsilon_d ; \; H_{33} =
H_l + U + 2 \epsilon_d
\label{diag}
\eneq
$H_{03} =H_{12} = 0 $, while the tunneling part of the Hamiltonian of 
eq.s (\ref{ofdia1},\ref{ofdia2}) provides the  remaining  
 off-diagonal terms $H_{ij}$  $(i,j =0,1,2,3)$.

  The effective Schr\"odinger equation
projected onto  $\Xi$  is: 
\beq
P [ ({\cal H}_0  - E ) - V^\dagger ( 1 - P ) 
( {\cal H}_0 - E )^{-1} V ] P | \Psi
\rangle = 0
\label{effeq}
\eneq
where $P$ is the corresponding projection operator.

For the sake of simplicity we shall take $v_{ 3 \alpha} ( q )$ and 
$v_{ \alpha 0} ( q )$ to be independent of $\alpha$. Then, 
to   second order in $v$, eq.(\ref{effeq} ) becomes (see Appendix A for details):

\begin{eqnarray}
H^{\rm \text{\it eff}} = H_l +\sum _{q q'} Q_{q q'} \sum _{\sigma}
 \phi_{ q \sigma}^\dagger \phi_{ q^{'} \sigma} +\nonumber\\
\sum _{q q'} J_{q q'} \sum _{a} \vec{S}_{\text{\it eff}}^a  
  \phi_{ q \lambda}^\dagger  \sigma ^a_{\lambda \lambda^{'}}
\phi_{ q^{'} \lambda ^{'} }
\label{hcace} 
\end{eqnarray}

where $\sigma _{\lambda \lambda '}^a$ are the spin $\met $ matrices 
$(a = x,y,z)$ and 
$\vec{S}_{\text{\it eff}} $ is a spin-$\met $ representation of the rotation 
within $\Xi $, which is well defined, provided the constraint 
$n_1 + n_2 = 1$ is satisfied.
 Indeed, in this case,    
$(\vec{S}_{\text{\it eff}})^2 $ = $\frac{3}{4} (n_1+n_2 -2n_1 n_2 ) =\frac{3}{4}$. 
More explicitly: $\Seff^z = \met ( d^\dagger _2 d_2 - d^\dagger _1 d_1 )$
and the ladder operator is $ \Seff^{-} =d^\dagger _1 d_2 $. 

The interaction matrix elements are
 
\begin{eqnarray}
Q_{q q'} =  -\frac{1}{2}   \left[ \frac{ v^*_{3 } ( q )
v_{3 } (q^{'} ) }{ \epsilon_d + U - \epsilon_q } + \frac{ v_{0 } ( q) 
v^*_{0 } ( q^{'}) } { \epsilon_d - \epsilon_{q^{'}}}\right ] \: , \nonumber\\
J_{q q'} =  2   \left[ \frac{ v^*_{3} ( q) 
v_{3} ( q^{'}) }{ \epsilon_d + U - \epsilon_q } - \frac{ v_{0} ( q ) 
v^*_{0} ( q^{'}) } { \epsilon_d - \epsilon_{q^{'}}}\right ] \: ,
\label{coupl}
\end{eqnarray}

corresponding to the potential scattering term  
and the Kondo coupling term, respectively.

The potential scattering term $Q_{q q'}$ provides a one-body interaction in the
``charge'' channel. It is irrelevant for the physics of the Kondo effect, which
is related to interactions in the ``spin'' channel \cite{tswieg},
 and we shall disregard it in the following.

By taking $q$ at the Fermi surface,  the Kondo term in eq.(\ref{hcace}) 
becomes:

\beq
H_{K} =H^{\rm  eff} - H_l \sim  J \vec{S}_{\rm
\text{\it eff}} \cdot \sum_{ q q^{'}}
  \phi_{ q \lambda}^\dagger  \vec{ \sigma}_{ \lambda \lambda^{'}}
 \phi_{  q^{'}\lambda^{'}} .
 \label{effint}
\eneq

Tuning  $V_g$ such that the chemical potential
of the leads  is  the zero of the single-particle energies, the  
Kondo coupling strength in eq.(\ref{coupl}) takes the usual form:
\beq
J = | V |^2 \left[ \frac{1}{ \epsilon_d + U } - \frac{ 1}{ \epsilon_d }
\right]
\label{effc}
\eneq
$\epsilon _d \approx -U/2 $ implies  $J >0$, i.e., 
the interaction between the spin density of the Fermi sea at the location
of the dot  and  the effective spin $\Seff$, described by  
eq.(\ref{effint}),  is  antiferromagnetic.
Thus, the low-energy physics of the system
will be controlled by the strongly coupled AF- $S=\met $-Kondo fixed point,
where  a spin singlet takes place at the impurity site.
 Because of the way $S$ is
related to $S_{ \rm eff}$ (eq.(\ref{phef} )), total compensation of 
$S_{\rm eff} $ leads to a spin $ S =\met $ at the QD.

In the next Section
we will make use of such an equivalence  to construct a formalism for 
the strongly coupled fixed point  of our system.

\section{Gutzwiller projection.}

In Section II we have shown that cotunneling   at the ST point  strongly
involves just two among the four available states of the isolated dot, which
 can be  described in terms of $\Seff$, including
both orbital and spin degrees of freedom locked together. Instead,  
 the other relevant degrees of freedom of the dot 
 can be lumped into another pseudospin 
variable $\vec{S}_r $  ($S_r =\met$) which  decouples  
below $T_K$,  because the dot is tuned at  CB   (eq.(\ref{phef})). 
According to 
 Section III  the  dynamics of correlations  involves the true spin density 
of the delocalized electrons at the dot site $\vec{\sigma}(0)$.  
Therefore,  the  study of the magnetization in the correlated state 
requires the knowledge
 of  $\vec{\sigma}(0)$.

In Section III we discuss the assumptions  under which 
  the  isotropic single channel spin $\met$ AF Kondo model  of 
eq.(\ref{effint}) describes the physics  at the 
 fixed point. The correlated state is  a
 Nozi\'eres local  Fermi Liquid (FL) \cite{noz},
 with  a  spin singlet at the origin. 
 In our case there is a substantial magnetic field $B\sim B_*$ 
and  a  small detuning $\delta B \neq 0$ is likely to occur.

 In this  Section,  we  construct
an approximated GS  of   the strongly  coupled  system, described by 
$H^{\text{\it eff}}$ (eq.(\ref{effint})). 
In order to describe the scaling of $\Seff$ toward total 
compensation, we use a variational method based on  the 
Gutzwiller projection (GP)
technique. This qualitatively reproduces  the  main  
features of the correlated singlet state. Detuning $B$ off
$B_*$ allows for probing the QD magnetization and for studying the magnetic
response in the correlated state. So, in Section V
we extend our approach  to the $B\neq B_*$-case.

As $J$ increases, states other than a 
 singlet at the impurity  become higher in energy  and  can be 
``projected out''  from the physical Hilbert space. This is quite
similar to what happens in the 1-d Hubbard model at large $U$, where
higher-energy states are ``projected out'' by the interaction 
and the GP method works successfully \cite{vol2}. Our approach is similar
in spirit to Yosida's variational technique \cite{yosida}. However, here we
are mostly interested in the ``macroscopic'' variable $\vec{\sigma} ( 0 )$.
Hence, it is the only lead operator involved in the construction of the 
variational state. This makes our technique much simpler than the one in 
\cite{yosida}, because our approach requires just one variational 
parameter for the trial state.

 Here we list the basic 
assumptions concerning  the model Hamiltonian and  the 
trial variational state. We introduce our approach for the simple
spin-$\met$ Kondo Hamiltonian (which corresponds to $B=B_*$). In Section V
we generalize it to the $\delta B \neq 0$-case.
To simplify  the notation, we drop the suffix $_{\text{\it {{eff}}}}$  
from $\Seff$ 
all throughout  this Section and the next one.

\begin{description}
\item[a.] 

The Hamiltonian is  (see eq.(\ref{effint})):
\beq
H = H_l + H_K = H_l + J a \vec{\sigma} (0)\cdot \vec{S}  
\label{modh}
\eneq
The impurity is located at $x=0$. The scattering is $s$-wave, so that 
the 
model is effectively 1-d and can be defined on a lattice with 
spacing $a$ and 
periodic boundary conditions at $L= N_l a$.
($N_l$ is the number of lattice sites ) \cite{notaeh}.
$H_K$ provides an effective interaction among  the 
delocalized electrons.    
At each stage of the scaling process, we  take  the  Slater determinant  
$| FS \rangle = \prod_\sigma \prod_{ q < 0 } c_{ q \sigma}^\dagger |
0 \rangle $  as a reference state (FS: Fermi sea).
 The $c_{q \sigma } $ operators annihilate 
the quasiparticles and  $|0\rangle$ is the quasiparticle vacuum. The spectrum 
is linearized around the Fermi point  $q_F$. $D$ is the bandwidth. 
Therefore, the lead  Hamiltonian is $ H_l =
 \sum_{ q \sigma } \epsilon_{ q \sigma } c_{ q \sigma}^\dagger c_{ q 
\sigma }$, where $ \epsilon_{ q \sigma } = v_F q$. Here  $v_F$ is the
Fermi velocity and $q$ is the momentum measured with respect to the Fermi
momentum. The bandwidth is $2D= 2\pi v_F /a$. 
The density of states at the Fermi level is 
$\nu ( 0 )= N_l/D$  and is assumed to be constant during the scaling.    

\item[b.] One can represent  the  real space field operator  in terms of 
the $c$-operators: 
\beq
\psi_\sigma (x) = \frac{1}{\sqrt{ L}} \sum _q e^{iqx} c_{q\sigma}.
\eneq

It satisfies the   anticommutation relations 
 $\{ \psi_{\sigma}^\dagger ( x ) , \psi_{\sigma^{'}} ( x^{'} ) \} =
\frac{1}{a} \delta_{\sigma ,\sigma^{'}} \delta_{x , x^{'}}$.  

 The quantity
$ an_\sigma ( 0 ) = a\psi_\sigma^\dagger ( 0 ) \psi_\sigma ( 0 ) $ plays 
the role of  
 the occupation number at the impurity site. This  is consistent
  with the operator identity $(
a n_\sigma ( 0 ))^2 =
a n_\sigma ( 0 )$. The corresponding  spin amplitude due to the delocalized 
electrons    at $x=0$  is
$a \vec{\sigma} ( 0 ) =  \frac{1}{2} a \psi_\alpha^\dagger ( 0 ) 
 \vec{\sigma}_{\alpha
\alpha^{'}}  \psi_{\alpha^{'}} ( 0 )$. 
The average value of $ ( a \vec{\sigma}
( 0 ))^2$ on $| FS \rangle$ is: $\langle FS | ( a \vec{ \sigma} (  0 ))^2
| FS \rangle$=3/8. This is a consequence of averaging over configurations 
with zero or double occupancy at $x=0$ (total spin 0) and configurations
with single occupancy at the same point (total spin 1/2).

\item[c.]
The correlation between the impurity and the FL 
 is accounted for by projecting out of the  uncorrelated state 
the  components with a triplet
or a doublet of the total spin at the impurity site,
  $\vec{S} + a \vec{\sigma (0)}$. 

Denoting  the projector as $P_g$, the variational trial state is defined as:
\beq
|g^\lambda \rangle = P_g |FS , \lambda \rangle \equiv  P_g \left ( 
|FS \rangle  \otimes  |\lambda \rangle \right ) \: . 
\label{trial1}
\eneq
Here $|\lambda \rangle $ is an eigenstate of $S^z $ with eigenvalue 
$ \met \lambda   = \pm \met$ and   $g$ is the variational parameter. 
The explicit form of the Gutzwiller projection operator $P_g$ 
will be given below. So far, it is enough to say that  all the components 
of the state other than a localized singlet at $x=0$ are projected out,
as  $g$  changes from zero to  the limiting value,  $g=4/3$.
 
\item[d.]  At $ B =B_*$,  we assume the usual  one parameter scaling
process  when  $D \to D -\delta D$.  $g$ approaches 4/3, and, 
in order to to guarantee that at each step the total energy is at a minimum, 
the parameter $\nu (0 ) J$ is correspondingly  renormalized.
 If $\delta B = B - B_* \neq 0$, a second parameter appears 
in the
model, related to $\delta B$. We shall refer to such a parameter as $h$
and will introduce it in the next Section. 
It  is generated by the shift in
the quasiparticle energies, according to their spin and will determine the
``off-critical'' magnetization (see point e).

\item[e.]  According to eq.(\ref{phef}), the effective  magnetization 
of the dot is given by the average of the total spin
at $x=0$  on the trial state $| g^\lambda \rangle $
  (shortly denoted by $\langle ...\rangle _{g,\delta B} $): 
\begin{eqnarray}
\lim_{g \to 4/3} \left \{ M_d (g,\delta B) - M_d (g,\delta B =0)
\right \}\nonumber\\
\approx   \lim_{g \to 4/3}  g^* \mu _B \left [    \langle  
\Seff^z \rangle _{g,\delta B } + \langle 
a \sigma ^z (0) \rangle _{g,\delta B} \right ] 
\nonumber\\
  \equiv   M_{imp} + m( h) \:\: .  
\label{mtot}
\end{eqnarray}
Here we have temporarily restored the suffix $ _{eff}$, for clarity.
The dot magnetization   $M_d$ includes 
the local  moment on the impurity site $  M_{imp} $, and 
the induced magnetic moment 
of the delocalized electrons from the contacts, $m(h)$.

At $\delta B =0$ both     $M_{imp} $ and $m(h) $ vanish,
  due to  the compensation  of
 the effective spin $\Seff $ of the standard spin $\met $ 
AF Kondo problem. $ M_d (g,\delta B =0)$ is the residual magnetization 
at $B= B_*$  associated to the expectation value of $S_r$ of eq.(\ref{phef}) 
and is not discussed in this paper. However, the spin susceptibility
 of $S_r$  is expected to be much smaller than the Kondo 
susceptibility  of $\Seff$. 

One remark concerns the fact that the giromagnetic 
factor in eq.(\ref{mtot}) might not be  the
bare one, appearing in the   electron Zeeman spin splitting.

In Section V we work out the form of  $M_{imp}$  and $m(h)$ 
in the strongly coupled scaling  regime.

\end{description}

Let us now construct $P_g$.

According to Noziere's FL picture of the correlated Kondo state
\cite{noz}, the main effect of the interaction between the FS and
the impurity taken at the AF fixed point 
 is to  fully project out  the components 
of the  trial state other than a singlet of the total spin 
at the impurity site.
 Let us consider the operator
\beq
a \vec{\sigma} ( 0  )\cdot \vec{S} = \met [(a \vec{\sigma} ( 0  ) + \vec{S})^2
-\frac{3}{4} - (a \vec{\sigma} ( 0  ))^2 ]
\eneq
It will give zero when acting on a state that has a doublet on the impurity 
site, (i.e. $\langle (a \sigma (0) )^2\rangle =0$),  $-3/4$ when acting on a singlet, $1/4$ when acting on a triplet. 
Among the three possibilities  listed above, the state corresponding to 
the triplet is the one highest in
energy. We fully exclude it from the trial state from the outset and gradually
project out the doublet  by varying the parameter $g$  from 0 to 4/3. 
Accordingly, we define $P_g$ as:

\begin{eqnarray}
P_g = \left( 1 - \frac{3}{4} g - g a \vec{ \sigma } ( 0   )\cdot \vec{S}
\right) ( 1 - 4 a \vec{ \sigma } ( 0 ) \cdot \vec{S} ) =
\nonumber\\
\left( 1 - \frac{ 3}{4} g + g ( a \vec{ \sigma } ( 0))^2 \right) -
4 a \vec{ \sigma } ( 0 ) \cdot \vec{S}
\label{pr3}
\end{eqnarray}
This operator projects out the components other than the localized singlet.
 When $g$=4/3 the projection  is complete. 

In this section we report the result for  the zero magnetic field 
case. The key points of the calculations are summarized in  Appendix  B.

The expectation value of the Hamiltonian in eq.(\ref{modh}) differs from the
energy  of the reference Fermi sea $|FS\rangle$, $E_0 = - 2 \pi v_F  N_l / a$,
 just
by a term of order $ O(1)$ in the particle number $N_l$. 
In units of $D$, this  correction   is:

\beq
\epsilon (\xi, j) = \frac{ 1}{D}\left ( \frac{ \langle g^\lambda | H | g^\lambda \rangle }{
N [ \xi ] }  - E_0 \right )  =  2\frac{   1 + \xi^2  - 
 \xi - j }{  1 + \xi^2 }
\label{ener2}
\eneq
where $N [ \xi ]$ is 
norm of the trial state (see eq.(\ref{norm})) and we have introduced the 
dimensionless variables $j= 3\nu(0)J/(8 N_l)$ 
and $ \xi = \frac{3}{8}(\frac{4}{3} -g) $ for convenience.   

It is worth stressing that $\epsilon ( \xi , j )$ does not depend 
on the polarization of the impurity, $\lambda$. Such a degeneracy 
disappears when $\delta B \neq 0$\cite{nota4}. 

For each $j \propto J/D$, eq.(\ref{ener2} ) takes a minimum 
as a function of $\xi$ (i.e. $g $).  
The  minimum   $\overline{\xi}( j)$ w.r.t. $j$ is given by:

\beq
 \overline{\xi}(j)  = -j +  \sqrt{ j^2 + 1}
\label{sol}
\eneq
Both the triplet and the doublet are fully projected
out of the trial GS   when $g \to 4/3$, i.e. when 
 $\overline{\xi} $ flows to zero. Eq.(\ref{sol}) shows that this limit 
corresponds to  $j \rightarrow \infty$.

The derivative  of  eq.(\ref{sol})  w.r.t. $\overline{\xi}$,  reproduces  
 the  poor man's scaling law  for $j$
\cite{poorman} provided $g$ is small (that is, $\xi$ is close to 1/2)
:
\beq
\frac{ d  j  }{ d (ln\xi ) } \approx -  j^2 .
\label{scaling}
\eneq
Eq.(\ref{scaling}) has been worked out in \cite{poorman} within 
perturbation theory, provided $g$ is small, that is, $\xi$ is close to
$\met$. In our 
simplified approach , eq.(\ref{sol}) (not eq.(\ref{scaling}) !)
 holds  all the way down to the fixed point.

\section{The magnetic moment}

In this Section we qualitatively discuss what happens to the energy 
and to  the magnetic 
moment of the dot when a magnetic field is present  and $B$  is  
possibly detuned off $B_*$.

Here we  do not consider the in detail the magnetic moment associated to
$S_r$. Instead, we focus on the one in the third 
line of eq.(\ref{mtot}).  We generalize the construction of
Sect IV to the $\delta B \neq 0$-case 
by shifting the energies of the quasiparticles of the FS
according to their spin.  Two corrections to the magnetic moment
arise due to this shift: $ a)$ $\langle \Seff \rangle_{g,B}$ 
may not scale down to 
zero, $ b)$ an induced  moment of the delocalized electrons 
 $m(h)$ may arise. We first discuss the latter contribution. 
In the absence of correlations,  because $B_*$ is the degeneracy point,
the energy difference of electrons involved in cotunneling 
at the Fermi level is the same for both spins.  Therefore, 
provided also the density of states of the conduction electrons is 
the same at the Fermi energy for  both spins, $B_*$ will only be
responsible for changes in the tunneling Hamiltonian  due to orbital 
effects \cite{nota5}, which do not influence  the flow to strong coupling 
in a relevant way. 
 
 This is not the case if $\delta B \neq 0$. Qualitatively, there will be 
changes in the number of the conduction electrons  involved,  
according to their spin  $\sigma = \pm \met $ as follows:
\beq
\frac{ N_\sigma  -N_l/2 }{N_l/2} = 
 \sigma \delta B \frac{\nu (0) }{N_l}=
\sigma \frac{ \delta B}{D}   
\label{mraw}
\eneq

The interaction  modifies eq.(\ref{mraw}) by scaling, so, we generalize
it by defining a dimensionless quantity $h$ in terms of the expectation value
of the quasiparticle spin at the origin evaluated on the reference state
$| FS \rangle$:

\begin{eqnarray}
\langle FS ,\lambda | a \sigma^z ( 0 ) | FS , \lambda
\rangle _{\delta B \neq 0} = \nonumber\\
 \langle FS , \lambda | \met \left (a n_\uparrow ( 0 )
- a n_\downarrow ( 0 ) \right )  | FS , \lambda
\rangle \equiv   \met h_\lambda    \: ,
\label{mag1}
\end{eqnarray}
($h_\lambda = \lambda \cdot h$, where  
$\lambda = \pm $ is the polarization of the impurity ).
Eq.(\ref{mag1}) states that  
 the magnetic field produces a spin density at the  impurity
site proportional to $h$. When  $\delta B = 0$, rotational symmetry of
the FS implies $h=0$ unless a local paramagnetism takes place,
 so that  $h$ is a homogeneous function of $\delta B$.

Eq.(\ref{mag1}) is enough to generalize eq.(\ref{ener2}) to the 
$\delta B \neq 0 $-case. Using the results of of Appendix B,
we  get:
\begin{eqnarray}
\epsilon _\lambda ( \xi , h, j ) = \nonumber\\
  \frac{ [ 2 ( 1 - \xi + \xi^2 )  +
h_\lambda ( 1 - 2 \xi ) ] ( 1 - h_\lambda ) - 2 j ( 1 +
h_\lambda )}{  [ ( 1 + h_\lambda ) + \xi^2 ( 1 - h_\lambda ) ] }
\label{enerh}
\end{eqnarray}

As in the previous Section, the minimum of eq.(\ref{enerh})
$w.r.t.\:\:\xi $ $( \partial \epsilon / \partial \xi =0 )$ 
fixes  $\overline{\xi}^\lambda ( j,  h) \sim (1+h)/2j $. 
The states $|g^\lambda \rangle  (\lambda =\pm )$ are no longer degenerate, 
except for $h=0$ and they represent the ground and excited state, depending 
on the sign of $h$ ( Fig.2b). At $h=0$ the  delocalized electrons do not 
have an intrinsic magnetic moment close to the fixed point:  $m(h=0)=0$. 
Because of the degeneracy, the magnetization
at $h \sim 0$ is obtained from the average  $E/J = ( \epsilon _+ +
\epsilon _- )/2 $.  Minimization with respect to $\xi $ yields 
the leading large $j$-correction to the energy for  $h \neq 0$, due to  
the delocalized electrons  of the leads:
\begin{eqnarray} 
\frac{E }{J} \to 
  \frac{3}{4} \left (-1  +\frac{1}{j_o} +  \frac{h^2 }{j_o}+
{\cal O} (h^4) \right ) \: .
\label{enerh1}
\end{eqnarray}
Hence, the contribution from the delocalized electrons to the 
magnetic moment  is  linear in $h$ as expected from the exact result
\cite{tswieg}. However, the relation between  $h$ and $\delta B$ 
cannot be established within this model, unless we state some direct link
between the strong coupling fixed point and the uncorrelated state 
previous to scaling. This  link  can be inferred from the  
educated guess  that    the response of the conduction 
electrons to  $\delta B$, given by  
eq.(\ref{mag1}) in the absence of correlations,  smoothly evolves into 
the one  of the correlated 
state, $ m(h) $,  in the form:
\begin{eqnarray}
m(h) =
 \lim _{j \to \infty} \langle g^\lambda | a \sigma ^z ( 0 )
|g^\lambda  \rangle _{j, h} \sim  \met    \frac{T_K}{J} h_\lambda 
\label{ansatz}
\end{eqnarray}
(here   $T_K$ is the Kondo temperature).

Indeed, by using the insight we have from  the exact solution, we infer that, 
the fraction of delocalized electrons involved in the correlated state 
is expected  to be ${\cal{O}}(T_K/J)$. Because $T_K $ is a scale invariant, 
both $h$ and $j$ scale to larger values in the same way as $\xi \to 0 $. 

Then, by comparing   
eq.(\ref{ansatz}), supplemented with the derivative of 
eq.(\ref{enerh1}), to the 
magnetization of the uncorrelated state, 
$\met  (N_\uparrow -N_\downarrow ) / N_l  =  g^*\mu _B \delta B  /4 D$,
we obtain: 
 \beq
\frac{1}{2J}\frac{\partial E}{\partial h}\sim  \mu_ Bg^* \frac{\delta B }{T_K}.
\eneq
Thus, we recover the expected  result that  the spin susceptibility is  
$\propto 1/T_K $\cite{tswieg}.

This concludes our discussion of  $m(h)$.

According to  eq.(\ref{mtot}),  the total magnetization also includes 
 the  term   $ M_{imp}$.
This term 
just provides a subleading correction to the total magnetization.

 Indeed, $ M_{imp}$  is given by 

\begin{equation}
 M_{ imp } ( \xi , h_\lambda ) = \frac{1}{4}  \mu_B g^*
 \lambda  \frac{ \xi^2 }{ \frac{ 1 + h_\lambda}{ 1 - h_\lambda } 
+ \xi^2 } \:\:  ,
\label{resmagn}
\end{equation}
i.e. this  correction to the fractional spin of the dot decreases as the 
second power of $\xi$ when $\xi $ decreases in the strong 
coupling  limit. 

\section{Concluding  remarks }

Kondo coupling  is the striking realization of 
non-perturbative tunneling between   a  QD in the CB region  and 
contacts. It requires the degeneracy of the GS dot level, what 
certainly takes place when $N$ is odd. In the even-$N$ case,  the 
GS can be  a triplet, because  of Hund's rule and corresponding 
partial screening of  the $S=1$ spin on the dot has been
 observed \cite{silvano,tarucha}.
The proximity to a Triplet-Singlet (T-S) transition 
enhances the coupling and increases the Kondo temperature\cite{eto}.
A small magnetic field driving the system across the level crossing
 can tune the  intensity of the effect  due to the T-S conversion. 
This enhanced Kondo  conductance  was indeed measured for  $N=6$. 

In \cite{noi,pustilnik} a different mechanism for the Kondo effect
in an even-$N$ QD has been proposed, considering states belonging to
different representations of the total spin, $S$=0,1, at the accidental
crossing between the singlet and the triplet state lowest in energy.
In \cite{pustilnik} $B$ is taken in the plane of the dot and the crossing 
is attributed to the Zeeman spin splitting term. Because the latter 
is usually very small, it is  unlikely that this can 
occur anywhere else but at the point mentioned above. Then the mechanism 
discussed  by   
\cite{silvano,eto} seems to be  more likely\cite{nota6}. 

On the contrary, by  increasing $B_\perp$ up to a suitable value 
 $B_*$,   the degeneracy of the triplet state is lifted and the singlet 
state ceases to be the GS of the system. In \cite{noi}  
we estimate quantitatively the parameter values of the optimal device
by using results of  exact diagonalization 
for few interacting electrons in a dot\cite{nota}.
 A vertical QD with a magnetic field along the $z-$axis, 
undergoes a S-T transition mainly due to 
the Zeeman term, which favors larger angular momenta  along $z$, 
and to the correlation. The spins of the electrons 
in the contacts are  still
 unpolarized at the Fermi energy if the contacts  are 
bulk  (i.e. 3-dimensional) metals. This allows Kondo coupling to occur
at the crossing between  the  two lowest levels among the four with $S=0,1$.
In \cite{pustilnik} the zero voltage anomaly in the differential conductance
is discussed by including just these two levels.
  The fact that two different spin states are involved is reflected  in 
the asymmetry in the transmission probabilities w.r.t. the spin index.
This  can be traced back to a term which couples the charge and spin 
degrees of freedom in the exchange Hamiltonian that can be derived from a 
Schrieffer-Wolff transformation.  However, 
this term is irrelevant as the system flows toward the  
strongly-coupled  fixed point, consistently with the full decoupling of 
the charge and the spin degrees of freedom in such a limit, due to the
widely different energy scales for the two excitations \cite{tswieg}. The 
  Schrieffer-Wolff transformation  (see  Section III)
 maps the problem onto the AF spin $\met -$ Kondo Hamiltonian. 
Coupling is better described  by splitting the total dot spin $S=0,1$ 
 into two spins $\met$, which we refer to as
$S_r$ and $\Seff$ (see eq.(\ref{phef})). 
$\Seff$ is totally screened, independently on $B_*$,  
while a residual spin $\met$ survives,  whose magnetic 
moment depends  on $B_*$. 

In this work we have concentrated on the magnetization at the strong coupling 
fixed point. A variational approach is used for the GS, based on
the Gutzwiller projection method, which, as far as we know,
 has never been  applied to the  Kondo problem.  
This technique  allows for studying a small detuning of $B$ off
$B_*$ and for  discussing the dependence of the 
magnetic moment   on $\delta B = B -B_*$.
The basic assumption is that  $\delta B \neq 0$  does not 
move the system away from the strongly coupled state. The scale of magnetic 
field  at which  the Kondo state is disrupted, is expected to be quite large 
(it was estimated in \cite{noi}  to be of the order of 1~Tesla). 
A similar   conjecture  in the case of magnetic  impurities in metals 
has been formulated by
Y. Ovchinnikov $et \; al.$ \cite{ovchinnikov}, starting from a 
perturbative approach in a reduced Hilbert space. 

When $\delta B \neq 0$, besides the Kondo coupling
parameter $j \propto \nu(0)J$, 
our variational 
correlated state includes a new  coupling parameter which is related to the 
detuning magnetic field, $h_\lambda \propto \delta B /T_K$. 
As the system flows toward strong coupling, this parameter increases 
together with  $j$.  Within our formalism we show that
 $j$ scales according to Anderson's poor man's scaling,
provided the ration $h/j$ keeps small ($<1 $).
 
   We have proved that the  magnetic moment on the dot has a term linear in 
$h$   (see eq.s(\ref{mtot},\ref{ansatz})),  
which arises from the screening due to  
the electron density of the Abrikosov-Suhl resonance located at the 
dot site. Still,  more work has to be done in order to  provide the  
relation between $h$ and  $\delta B$ (in our formalism $h$ is introduced 
as the  shift of the  quasiparticle energies corresponding  to  
$\nu (0) \delta B$).

 The QD GS includes  what we referred to as a spinon, 
carrying a  magnetic moment $S_r =1/2 $ but no  charge. The splitting 
in energy with magnetic field arising from this term
will depend on the spinon  wavefunction itself. 
At this stage, an estimate of the  renormalized  giromagnetic  factor
is impossible.
However,  the residual half  integer  spin on the dot, 
 together with the screening effects in the GS at non zero $\delta B$,
  has important  experimental implications.

In \cite{noi} $T_K$ has been estimated to be rather low, i.e., of the order
of tens of mK. At such a low temperature a transport measurement is 
cumbersome. We  proposed to probe the fractional magnetization of
the QD by means of a magnetic resonance experiment with microwaves.
 The energy splitting of the 
dot magnetic moment   can be detected by observing resonant absorption
as $B$ is slightly detuned from $B_*$.
 
\begin{figure}
\centering \includegraphics*[width=0.7\linewidth]{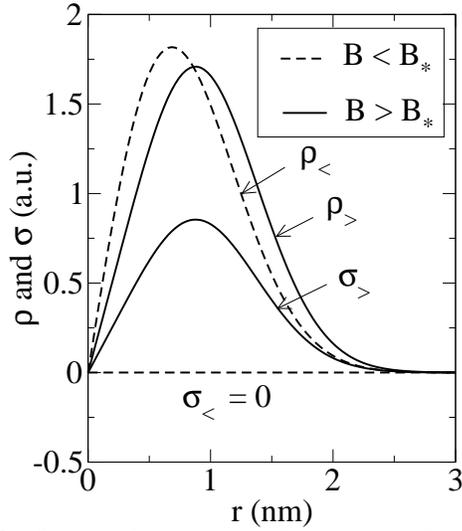}
\caption{Charge density as a function of the radius $r$. $B<B_*$ ($B>B_*$) 
should be understood as $B$ slightly smaller (higher) than $B_*$. 
The charge density $\rho$ is only slightly affected when $B$ goes across $B_*$.
The spin density $\sigma$ is zero when $B<B_*$. It has  the profile
of the charge density as $B>B_*$ ($\sigma _> = \met \rho _> $).}
\label{radial}
\end{figure}

\begin{figure}
\includegraphics*[width=0.8\linewidth]{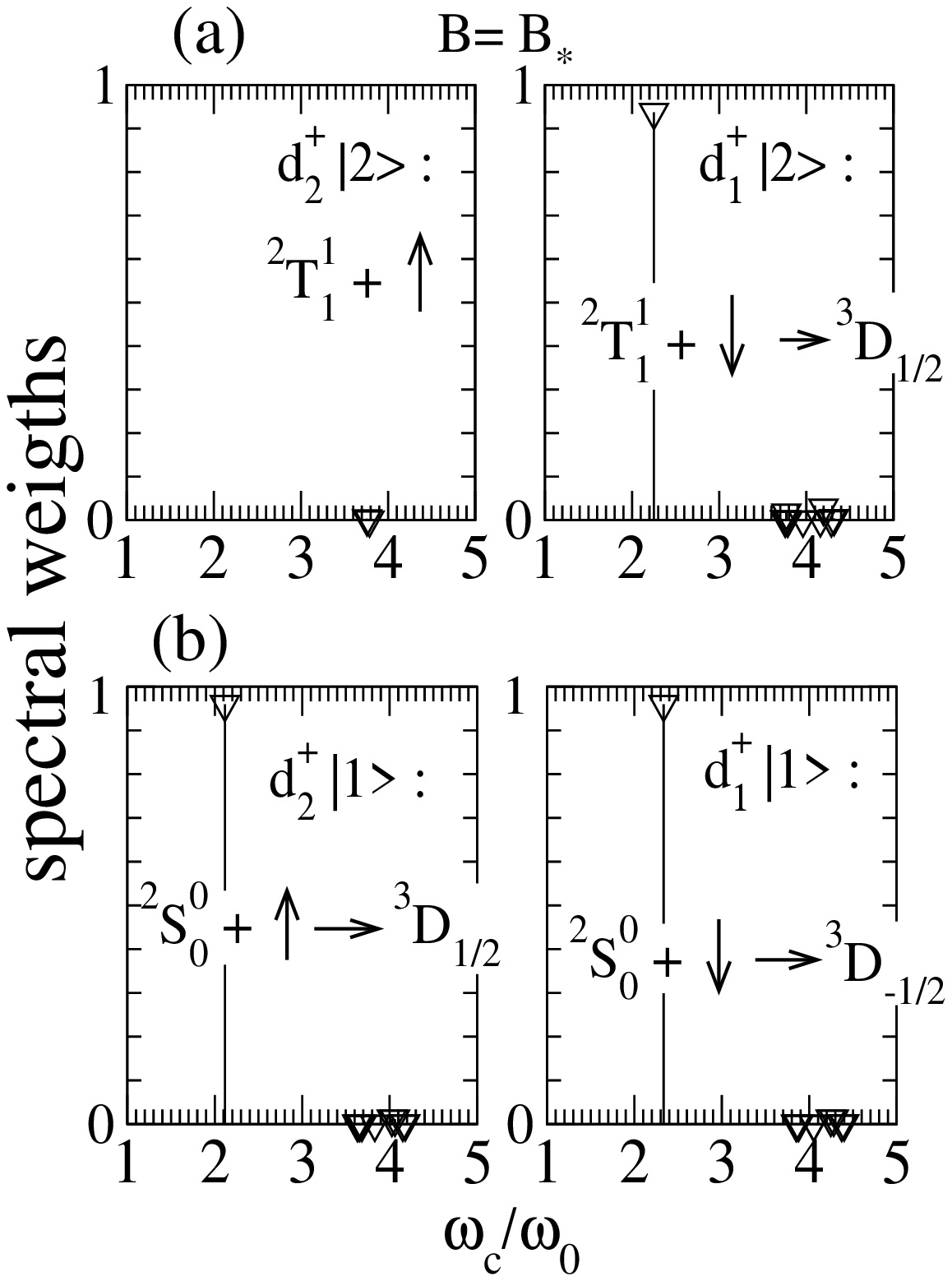}
\caption{Spectral weights for addition of a spin up or down to the N=2 states.
 (a) Top panel-left: Spectral weight for addition of a 
$(m=1,\uparrow)$-spin electron to
$~^2 T_1^1$; -right: Spectral weight for addition of a
 $(m=0, \downarrow)$-spin 
electron.
 (b) Bottom panel-left: Spectral weight for addition of a
 $(m=1,\uparrow)$-spin 
electron to $~^2 S_0^0$; -right:  Spectral weight for addition of a 
$(m=0, \downarrow)$-spin electron. Separation of the peaks is due to the 
Zeeman spin splitting}
\label{weight}
\end{figure}

A Knight shift should be measured, depending on the sign of $\delta B$,
which is  a consequence of the term $m(h)$ in eq.(\ref{mtot}).
 In Fig.~\ref{radial}  we plot the 
radial behavior of the $z-$ component of the total spin in the GS
when  $B$ is slightly 
less or 
larger than $B_*$. The 
dot is  isolated and   in the Coulomb blockade regime at $N=2$.
  In absence of Kondo coupling, screening from the 
delocalized electrons of the contacts does not occur and the 
magnetic moment changes abruptly from zero to  its maximum value at 
the transition. On the contrary, in the Kondo regime,  the spin 
susceptibility of the conduction electrons on the dot is expected to be large 
and a temperature dependent screening should be present  in the 
neighborhood of $B_*$. This   gives rise to a smooth  
 crossover in the paramagnetic resonance close to $B_*$ at low temperature. 

In conclusion, we have discussed the new possibility that  Kondo effect
arises in a QD with $N=2$ at low temperature, when the dot is tuned to
the ST point and the coupling  to the leads
increases. This happens when the GS  of the unperturbed dot 
changes from $^2S^0_0$ ($B < B_*$) 
to  $^2T^1_1$ ($B > B_*$), due to level crossing. 

The theory of \cite{noi} and of this work is not applicable to the 
TS crossing. Orbital effects are dominant in our case in  producing the 
crossing  as well as the properties of the many body states with $N+1$ or 
$N-1$ particles, as described in the text. All our results about the nature 
of the states involved are totally independent on the Zss. However, at the 
ST point  the Zeeman spin splitting is quite substantial, and favors a 
certain  component of $S^z$. Its role  becomes  important  in two respects: 

a)  Only two $N$-electron levels cross at the TS point 
(not four oh them, as is the case of the  TS point), which we label by 1 and 2.

b)  states at $N$, $N+1$ and $N-1$, although they are many-body levels,
have quantum numbers such  that   
there is just one  channel by which an electron can be 
virtually added or  subtracted  to the dot  in the $N$-state, because
 spin and orbital momentum of the  extra electron are strictly "locked 
together" in the selection rules. 

The spin of the QD  becomes $\met$ without changing the average occupancy 
\cite{noi}. We propose to probe spin fractionalization 
with   a magnetic
resonance experiment. A continuous shift in the resonance frequency 
should take place across the 
transition point between the two degenerate states. This is 
at odds with  an abrupt onset of energy absorption  at $B > B_*$
that is expected in the absence of the Kondo coupling.

\appendix

\section{The Schrieffer-Wolff tranformation.}

In this Appendix we report the details of the  Schrieffer-Wolff transformation,
leading to the Effective Kondo-like Hamiltonian (\ref{hcace}).

The Hamiltonian matrix elements $H_{ij}$ between states 
$|i \rangle, |j\rangle ( i,j =0,1,2,3 )$
 defined in and below eq.(\ref{doth}) are  
introduced in and below eq.(\ref{diag}). The  Hamiltonian operator
(\ref{effeq}) projected onto   the subspace $\Xi$ generated
by the two degenerate levels ($\alpha , \beta =1,2 )$ is:

\beq
H^{eff} = ( H_{l} + \epsilon_d ) + 
\sum _{\alpha \beta} {\cal M}_{\alpha \beta} 
\eneq

The operators ${\cal M}_{\alpha \beta}$ are given  by:

\beq
{\cal M}_{ \alpha \beta } = H_{\alpha  0} \frac{ 1}{ E - H_l} H_{ 0 \beta} + 
H_{ \alpha 3} \frac{ 1}{ E - U - 2 \epsilon_d - H_l} H_{ 3\beta}
\label{elem}
\eneq

The basic approximations we introduce in order to compute the 
${\cal M}_{\alpha \beta}$ consists in keeping only lead excitations 
with energy around  the Fermi level, whose energy is   negligible. 
Moreover, we approximate
$E$ at the denominators with $\epsilon_d$. Thus, the matrix elements become:

\beq
{\cal M}_{ \alpha \beta } = \frac{ H_{ \alpha 0} H_{ 0\beta}}{ \epsilon_d } - 
\frac{ H_{ \alpha 3} H_{3\beta }}{ U + \epsilon_d } 
\eneq

Their explicit form is:
\begin{eqnarray}
{\cal M}_{ \alpha \alpha } = 
\frac{1}{ \epsilon_d }  d_\alpha^\dagger d_\alpha
 \sum_{ q q^{'}} v_{ \alpha 0}^* ( q )
v_{ \alpha0} ( q^{'} ) \phi_{ q \sa} \phi^\dagger _{ q^{'} \sa }
\nonumber\\
- \frac{1}{ U + \epsilon_d }   d_{\ab}  d_{\ab} ^\dagger 
\sum_{ q q^{'}} v_{ \alpha 3}^* ( q ) v_{ \alpha 3} ( q^{'} )
\phi_{ q \sab}^\dagger \phi_{ q^{'} \sab} 
\nonumber
\end{eqnarray}
\beq
{\cal M}_{\alpha \ab} =  d_\alpha ^\dagger d_{\ab} 
\sum_{ q q^{'}} \biggl[  \frac{ v_{\alpha 3}^* ( q ) v_{\ab 3} 
( q' ) }{ U + \epsilon_d } -
\frac{ v_{\alpha 0} ( q' ) v_{ \ab 0}^* ( q ) }{ 
 \epsilon_d} \biggr]\phi^\dagger_{ q \sab} 
\phi_{ q^{'} \sa }
 \nonumber 
\eneq

Here is $\ab \neq \alpha $ and $\sigma _{1 (2) } = \downarrow (\uparrow ) $.

The Effective Hamiltonian can be expressed as the following operator:

\[
H^{eff} = H_l + \epsilon _d + \sum _q [A(q) + B(q) S^z ] +
\]
\[
 \sum_{ q q^{'}} U_c (q,q')   [ \phi_{ q \uparrow}^\dagger 
\phi_{ q^{'} \uparrow} + \phi_{ q \downarrow}^\dagger \phi_{ q ^{'} 
\downarrow } ] +\] 
\[ \sum_{ q q^{'}} U_s (q,q') \met  [ \phi_{ q \uparrow}^\dagger 
\phi_{ q^{'} \uparrow} - \phi_{ q \downarrow}^\dagger \phi_{ q ^{'} 
\downarrow } ] +
\]

\[ \sum_{ q q^{'}} {\cal J}' (q,q')   [ \phi_{ q \uparrow}^\dagger 
\phi_{ q^{'} \uparrow} + \phi_{ q \downarrow}^\dagger \phi_{ q ^{'} 
\downarrow } ] S^z +
\]

\[  \sum_{ q q^{'}} \biggl \{ {\cal J}^z (q,q') \met  
[ \phi_{ q \uparrow}^\dagger 
\phi_{ q^{'} \uparrow} - \phi_{ q \downarrow}^\dagger \phi_{ q ^{'} 
\downarrow } ] S^z +
\]

\beq
{\cal J}_\perp (q,q')   [ S^- \phi_{ q \uparrow}^\dagger 
\phi_{ q^{'} \downarrow} + \phi_{ q \downarrow}^\dagger \phi_{ q ^{'} 
\uparrow } S^+ ] \biggr\}
\label{bigboy}
\eneq
 
$\vec{S} \equiv \vec{\Seff} $ is defined in the body of the 
paper, after eq.(\ref{hcace}).
An enormous exemplification happens in eq.(\ref{bigboy}) if one takes 
the amplitudes $v$ to be independent on $q$ and real. If this is the
case, the coefficients are:

 \beq
A = \frac{ 1}{ 2} \frac{ v_{10}^2 + v_{ 20}^2 }{ \epsilon_d}   ; 
 B =  \frac{ v_{ 20}^2 - v_{ 10}^2 }{ \epsilon_d } 
\nonumber
\eneq
 \[
U_c = - \frac{1}{4} \left[ \frac{ v_{23}^2 + v_{13}^2 }{ U + \epsilon_d}
+ \frac{ v_{10}^2 + v_{20}^2 }{ \epsilon_d} \right]
\]
 \[
U_s = - \frac{1}{2} \left[ \frac{ v_{13}^2 - v_{23}^2}{ U + \epsilon_d} 
+ \frac{ v_{20}^2 - v_{10}^2}{ \epsilon_d}  \right]
\]
 \[
 { \cal J}^z = \left[ \frac{ v_{13}^2 + v_{ 23}^2 }{ U + \epsilon_d} -
\frac{ v_{10}^2 + v_{20}^2}{ \epsilon_d} \right]
\]
 \[
{ \cal J}_\perp = 2 \left[ \frac{ v_{13} v_{ 23} }{ U + \epsilon_d} -
\frac{ v_{10} v_{20} }{ \epsilon_d} \right]
\]
 \[
 {\cal J}^{'} =  \left[ \frac{ v_{13}^2 - v_{ 23}^2 }{ U + \epsilon_d} -
\frac{ v_{10}^2 - v_{20}^2}{ \epsilon_d} \right]
\]

and the Hamiltonian finally becomes:

\[
H^{eff} = H_l+ \epsilon_d  +A + B S^z + U_c  \rho (0) 
   + U_s   \sigma  (0) \]

\beq
+  \left \{ { \cal J}^z  \sigma^z(0) S^z  + 
{\cal J}_\perp  ( \sigma ^x(0)  S^x + \sigma ^y(0) S^y ) 
 {\cal J}' \rho (0) S^z \right \}
\eneq

where $ \rho (0) = \sum_{q q^{'}} \phi_{q \sigma}^\dagger
\phi_{ q^{'} \sigma} $  and $\vec{\sigma} (0) = \met 
\sum_{q q^{'}} \phi_{q \alpha}^\dagger
\vec{\sigma }_{\alpha\beta} \phi_{ q^{'} \beta } $

The extra terms in the Hamiltonian are:
$i)$ a  renormalization of the relative position of the two degenerate levels.
(this can be eliminated by re-tuning $B$) ;
$ ii)$ a potential scattering term in the charge channel;
$iii)$  a potential scattering term in the spin channel;
$iv$) an anisotropic Kondo coupling;
and, finally,  $v)$ a  spin-charge coupling.

If the  couplings to the states 1 and 2 are symmetrical, 
$v_{\alpha 0} $  and  $v_{3\alpha } $ do not depend on $\alpha$,
the terms $iii)$ and $v)$ are zero and the model 
becomes isotropic ( ${\cal J}^z = {\cal J}_\perp $). This leads 
to eq.(\ref{hcace})  and to the matrix elements of eq.(\ref{coupl}). 

Nevertheless, even in the non symmetrical case the potential scattering 
terms do not matter for the Kondo physics anyway, because  they can be  
re-absorbed in a redefinition of the energy
levels of the leads. Moreover, a ``poor-man'''s scaling argument shows 
that the coupling ${\cal J}^{'}$  does not scale as one lowers the energy 
cutoff (as confirmed by the corresponding differential equation in 
\cite{pustilnik}, as well). Hence, one can safely neglect it, as we
have done in the paper. Finally, in the 
range of the parameters relevant for our system, the fixed point
of the unisotropic Kondo Hamiltonian we have obtained is the same as the fixed
point of its  isotropic limit, so that our model Hamiltonian (\ref{hcace})
describes all the relevant physics of the problem.

The results we obtained with the Schrieffer-Wolff transformation
basically agree with the ones in \cite{pustilnik}. However, at odds with 
\cite{pustilnik}, in working out the transformation we choose to keep the 
couplings to  the $N=1$  and the $N=3$ levels different. Indeed, both should 
be taken into account, unless one makes a special choice of $V_g$, which 
appears to be unjustified \cite{nota7}.

\section{Relevant expectation values.} 

In this appendix we review the basic rules for the calculations leading 
to the results of Section IV and Section V.

We start by computing the norm of the state $| g^\lambda \rangle $ 
defined in eq.(\ref{trial1}),with the projector $P_g$ given by 
eq.(\ref{pr3}).
\begin{eqnarray}
N [ g ] = \langle g^\lambda | g^\lambda \rangle
= \langle FS , \lambda | ( P_g )^2 |  FS , \lambda \rangle 
\nonumber\\
= 2 ( 1 + \xi^2 )
\label{norm}
\end{eqnarray}
where  $\xi = \frac{ 1}{ 2 } \left( \frac{1}{ 2} - \frac{ 3}{4} g
\right)$, as always.
In order to reduce higher powers of the angular momentum in $(P_g)^2$ we
make use of  the general property of the  spin-1/2 matrices,
$S^a S^b = \frac{ \delta^{ab}}{4} + \frac{ i}{ 2} \epsilon^{ abc} S^c $
and of the following identities:
\begin{eqnarray}
a \sigma^z ( 0 ) = \frac{ 1}{ 2} ( a n_\uparrow ( 0 ) - a n_\downarrow ( 0
))
\nonumber\\
( a \vec{ \sigma } ( 0 ) )^2 = \frac{ 3}{4} ( a n_\uparrow ( 0 ) + a
n_\downarrow ( 0 ) - 2 a^2  n_\uparrow ( 0 ) n_\downarrow ( 0 ) )\nonumber\\
 ( a \vec{\sigma } ( 0 ) )^4 = \frac{3}{4} ( a \vec{\sigma } ( 0 ) )^2
\label{acinque}
\end{eqnarray}
Using eq.(\ref{acinque} ) we get the final result
\beq
( P_g )^2 = ( 2 \xi )^2 +  \frac{ 16}{3} \left( 1 -   \xi^2 \right) 
( a \vec{ \sigma } ( 0 ) )^2  - 8 a \vec{ \sigma } ( 0 ) \cdot
\vec{ S}
\label{xi1}
\eneq 
Taking the expectation value of $\vec{S}$ on $| \lambda \rangle$
and  because $\langle FS | a n_\alpha ( 0 ) | FS \rangle = \frac{1}{2}$,
(where $  a n_\alpha ( 0 ) = \frac{1}{N}\sum_{ k k^{'}}  c_{
k \alpha }^\dagger c_{ k' \alpha } $) we 
use
\beq
\langle FS | a \sigma^z ( 0 ) | FS \rangle = 0 ; \;\; \; 
\langle FS | ( a \vec{\sigma } ( 0 ) )^2 | FS \rangle = \frac{ 3}{8}
\label{fina}
\eneq
to  obtain eq.(\ref{norm}).

Next, we calculate the expectation value of the Kondo interaction 
Hamiltonian $H_K$  defined in eq.(\ref{effint}).  Because 
$H_K$   and $P_g$  commute, we get:
\beq
E_K \cdot   N [ \xi ]  =
J  \langle FS , \lambda | P_g^2 a \vec{ \sigma } ( 0 ) \cdot \vec{S} |
 FS , \lambda \rangle  = -  \frac{ 3}{2} J 
\label{ap5}
\eneq
The  interaction between the delocalized conduction electrons and the impurity
provides corrections to the ground state  energy of the FS, $ 
\langle FS| H_l |FS\rangle = -E_{GS} =- N_l D$,  that is again of order 
${\cal {O}}(1)$
 in the  number of particles. Such a 
correction is expressed as:
\begin{eqnarray}
\Delta E _c  \cdot   N [ \xi ] =  \langle FS , \lambda | P_g [ H_l , P_g ] 
| FS , \lambda  \rangle = \nonumber\\
 \frac{ 4}{3} ( 1 - 2 \xi ) \langle FS , \lambda | P_g [ H_l , ( a
\vec{ \sigma }  ( 0 ))^2 ] | FS ,\lambda \rangle \nonumber\\
 - 4  \langle FS , \lambda | P_g [ H_l , a \vec{ \sigma } ( 0 ) \cdot 
\vec{ S}] | FS ,\lambda \rangle 
\label{kine1}
\end{eqnarray}

We now calculate  the r.h.s of eq.(\ref{kine1}). Because 
$[ H_l, a n_\alpha ( 0 ) ] = \frac{ 1}{ N} \sum_{ q q^{'}} \epsilon_q ( c_{ q
\alpha}^\dagger c_{ q^{'} \alpha } - c_{ q^{'} \alpha}^\dagger c_{ q \alpha
}) $,  we find
$\langle FS , \lambda | [ H_l , a n_\alpha ( 0 ) ] | FS , \lambda \rangle =
0$
and, using Wick theorem:
\beq
\langle FS , \lambda | a n_\alpha ( 0 ) [ H_l , a n_\alpha ( 0 ) ] | FS ,
\lambda \rangle = \frac{ E_{ GS}}{ 2 N_l} \: .
\label{red2}
\eneq
Next, one more contribution  to the kinetic energy is:
\begin{eqnarray}
 \langle FS  , \lambda | a \sigma^a ( 0 ) [ H_l , a \sigma^b ( 0 ) ] | FS
, \lambda \rangle =\nonumber\\
 \frac{ E_{GS}}{ 2 N_l} { \rm Tr} \left[ \frac{
\sigma^a}{2} \frac{ \sigma^b}{ 2} \right] 
= 3 \frac{ E_{GS}}{ 2 N_l} \nonumber
\end{eqnarray}
Finally, the variational estimate of the energy correction due to the 
coupling to the impurity,  which is ${\cal {O}}(1)$
 in the  particle number, is:
\beq
\epsilon (\xi, J) = \frac{1}{D}( \Delta E_c + E_K ) =
 \frac{  2 ( 1 - \xi + \xi^2 ) - \frac{ 3}{4} J/D }{  ( 1
+ \xi^2 )}
\label{funct}
\eneq
Because  $J/D = \nu (0) J/N_l $, we obtain eq.(\ref{ener2}) of the text.

In Section V we extend eq.(\ref{funct} ) to the $\delta B \neq 0$-case.
The new quantity that appears in the problem is $h_\lambda$, defined 
by eq.(\ref{mag1}).
 All the operator identities we have proved  so far  still hold in the
$\delta B \neq 0$-case.
However, being $h_\lambda \neq 0$, the average values of $ ( a \vec{\sigma }
( 0 ) )^2 $ is now given by:
\beq
\langle FS , \delta B | ( a \vec{ \sigma } ( 0 ) )^2 | FS , \delta B \rangle =
\frac{ 3}{ 8} \left[ 1 + ( h_\lambda )^2 \right]
\label{magA2}
\eneq
which we used in order to obtain $
\epsilon [ \xi , h_\lambda ]$ in eq.(\ref{enerh}).

Finally,
 the average value of the $z$-component of the
 $\Seff^z$ at nonzero $h$, which is required in eq.(\ref{resmagn}) 
is given by:
\begin{eqnarray}
\langle S^z_{ eff} \rangle = \frac{1}{N[\xi,h_\lambda ]} 
\langle FS , \lambda | 
P_g S^z P_g|FS, \lambda \rangle_{ \delta B \neq 0} =\nonumber\\
\frac{ \xi^2 ( 1 - h_\lambda^2 ) }{ 4 [ ( 1 + h_\lambda)^2 + \xi^2
( 1 - h_\lambda^2 ) ]}\:\:
\label{unnor}
\end{eqnarray}

All other calculations in the text are straightforward. 

\vspace*{0.5cm}

 We  acknowledge interesting discussions with C. Marcus,  D. M. Zumbuhl
 (D.G.) and Y.Nazarov (A.T.).

Work supported by INFM (Pra97-QTMD )
and by EEC with TMR project, contract FMRX-CT98-0180.

\begin{thebibliography}{99}
\bibitem{nota2} 
a similar  construction appears in \cite{pg} for the TS crossing point 
but there  the role of the two fictious spins is quite different from the 
here.
\bibitem{nota3}
a  model has been proposed for the case when $B$ is parallel to the dot 
plane\cite{pustilnik} which shows some similarities to ours at first 
sight. However  
 it does not apply to the ST crossing as is discussed in Section VI.
\bibitem{notaeh} For simplicity one electron per site will be implicitely 
assumed but this choice is inessential.
\bibitem{nota4} lifting of the degeneracy of the two variational states
when electron-hole symmetry is broken is a feature in common with 
\cite{yosida}.
\bibitem{nota5}  it would not be so in Yosida's  original approach.
\bibitem{nota7} Our result fixes  a sign mistake in  ref.\cite{pustilnik}.
where is immediately apparent that  the effective Hamiltonian, as it stands,
is not spin-rotationally invariant  in the limit in which couplings to states 
1 and 2 are equal, thus leading to an unphysical result.
\end{thebibliography}
\end{document}